# Polarization description of successive ferroelectric switching in hafnia

Guo-Dong Zhao,[1] Xingen Liu,[2] Zhongshan Xu,[1] Wei Ren,[3] Xiaona Zhu,[1,4]* David Wei Zhang,[1,5] and Shaofeng Yu.[1,5]

[1] *School of Microelectronics, Fudan University, Shanghai 200433, China*
[2] *School of Mathematical Information, Shaoxing University, Shaoxing 312000, China*
[3] *Physics Department, Shanghai Key Laboratory of High Temperature Superconductors, State Key Laboratory of Advanced Special Steel, International Centre of Quantum and Molecular Structures, Shanghai University, Shanghai 200444, China*
[4] *Jiashan Fudan Institute, Jiashan 314100, China*
[5] *National Integrated Circuit Innovation Center, Shanghai 201204, China*

Intertwined ionic conduction and ferroelectric (FE) switching in $HfO_2$ lead to extensive focuses. To describe its fundamental phenomena, we present a free-energy model describing the potential of ferroelectrics with successive FE switching paths, and extend the domain model of ionic conduction to ferroelectric domains. Associate theoretical analyses and first-principles calculations suggest a nesting-domain pattern with opposite piezoelectric loops during the nucleation-and-growth process in displacive FE-$HfO_2$. A collective oxygen ion conduction mechanism is also proposed with a field-dependent ionic conductivity following the Merz's law. We conclude that the ionic conductibility is concomitant with the ferroelectricity in $HfO_2$, and it may provide a new venue for pursuing low temperature fast oxide-ion conductors and artificial synapses.

*Introduction.*—Ferroelectrics is a type of noncentrosymmetric dielectrics with spontaneous polarization, which is reversable by external electric-field *E*. Some of ferroelectrics are known as fast ionic conductors. The type-I is ionic conducting at high-temperature phase and ferroelectric (FE) at low-temperature phase, where the same ions are responsible for both ferroelectricity and ionic conductibility, such as silver iodide tetratungstate [1-4]; the type-II can have coexisting ferroelectric and ionic conducting in the same phase, while different ions are responsible for two properties, such as $KTiOPO_4$ [5-10] and $Na_{0.5}Bi_{0.5}TiO_3$ [11]; the type-III FE ionic conductor $CuInP_2S_6$ (CIPS) [12-15] is recently revealed, where the Cu-ion is proposed migrating in an asynchronous scheme [16], and being responsible for both ferroelectricity and ionic conduction coexisting in the same phase.

FE ionic conductors lead to interesting research topics like theoretical model including both properties [17], description for polarization evolving during conductions [16], abnormal piezoelectric loops [13], the coupling between ferroelectricity and ionic conduction [8,10], and ionic control over ferroelectricity [18], etc. The new concept of "ferroionic" electrochemical-ferroelectric states [19,20] can also be relevant. Originating from the displacements of the same kind of ion, ferroelectricity and ionic conduction are conceivably deeply intertwined in type-III FE ionic conductors. Relevant researches are important for the understanding or achieving anomalous FE behaviors, multistate polarization states, novel way of ferroelectricity control, and ionic-mediated memristor synapses, etc., for advanced information storage applications [13,14,16,18]. **However**, there is a lack of satisfying free-energy model with polarization description for FE ionic conductors. **Moreover**, similar theoretical research is rare for the extensively focused ferroelectric $HfO_2$ (FE-$HfO_2$) [21], which has been considered a promising candidate base material for FE devices [22]. It should belong to the type-III mediated only by oxygen ions according to the recent reports [23-25], meanwhile it is rather of displacive than order-disorder (CIPS) type. There may be new phenomena in FE-$HfO_2$, of both academical and industrial interests.

Ionic conduction is known playing a vital role in the hysteresis behavior of $HfO_2$ [25,26]. The measured remnant polarizations ($P_r$) of orthorhombic and rhombohedral hafnia films show abnormal positive temperature dependence [25,27,28]. The polarization estimated from atomic displacements is also confirmed much smaller than that extracted from polarization-voltage (P-V) loops [25].

In this Letter, we present a free-energy model for FE ionic conductors with polarization description and periodicity, and a FE-extension for domain model describing ion diffusion in displacive FE-$HfO_2$. We investigate the associate successive polarization evolving during nucleation-and-growth in FE-$HfO_2$ from first-principles. Based on our study, the quantized charge transport is declared in ferroelectrics, the observations of anomalous nesting-domain with opposite local piezoelectric phase loops in FE-$HfO_2$ are explained from a new perspective, and a distinctive field-dependent ionic conductivity following the Merz's law is also introduced from the research area of ferroelectricity.

*Free-energy model.*—The theoretical free-energy model of FE ionic conductor is intricate and rarely reported. It is believed [16] that firstly by Scott in 1999 [17], such a model was given by modifying the Landau-Devonshire free-energy description, written as

$$\mathcal{F}(x,T) = A(T - T_C)x^2 + Bx^4 + C(T - T_I)x^6, \qquad (1)$$

where $x$ denotes the ion displacement, $T_C$ the FE phase transition temperature, and $T_I$ the phase transition temperature for ionic conduction. This model allows the parameter of the sixth-order to change sign at $T_I$, therefore $x$ can be infinite (ionic conduction) to lower the Landau free energy. However, there is no analytical relation between expected polarization <$P$> and <$x$>, and the energy suffers an unphysically fast decreasing above $T_I$ when $x$ is increasing without the driving force of $E$.

To properly build a free-energy model with polarization description, it is necessary to give a quick review for the rigorous polarization definition in the modern theory of polarization (MTP) [29]. As a bulk property, the (change of) electric polarization ***P*** is defined as an adiabatic current flow within materials. The dimensionless time $\lambda$ acts as its coordinate, and the changing from -1 to 1 denotes a full FE switching. ***P*** is built based on Berry phase or equivalently on Wannier centers, which are both built from Bloch wavefunctions. ***P*** should uniquely determine the periodic charge density $\rho(\mathbf{r})$, while in turn the phase information will be lost from considering $\rho(\mathbf{r})$ alone. With the phase uncertainty, bulk property ***P*** is a multivalued vector quantity with branches separated by the polarization quantum ***Q*** = $e\mathbf{R}/\Omega$, where $e$ is the electron charge, ***R*** the lattice vector, and $\Omega$ the u.c. volume. Therefore, for traditional double-well ferroelectrics with finite FE distortion, a meaningful ***P*** is only well-defined within the center-branch, i.e., modulo (*mod*) ***Q***.

On the other hand, for FE ionic conductors, we must consider explicitly the value of phase twists or lattice vector shifts of Wannier centers, i.e., the value of quantized charge transport firstly applied in ferroelectrics [29,30]. As reported by Neumayer et al. [13] and O'Hara et al. [16], the inter-u.c. traveling of conducting ions (successive FE switching) enables ***P*** being infinite in the type-III FE ionic conductor CIPS. The conduction of an active ion in CIPS or FE-HfO$_2$ can be viewed as a successive FE switching, via multiple head-to-tail connected FE switching paths. Such paths can be sterically three-dimensional while topologically only one-dimensional to allow the long-range conduction of ions, as proposed by Habbal and Scott [17,31].

Based on the above introductions, the free-energy model must be periodic for describing successive FE switching. Taking scalar polarizations for simplicity, a trial model is written in the form of trigonometric series expansion, as

$$\mathcal{F}(\lambda, T, E) = A_0 + (T_C - T) \sum_{n=1}^{\infty} A_n \cos\left(\frac{n\pi}{m}\lambda\right) - EP(\lambda)$$

$$\approx A_0 + (T_C - T) \sum_{n=1}^{m} A_n \cos\left(\frac{n\pi}{m}\lambda\right) - EP(\lambda), \qquad (2)$$

where $E$ is the external electric-field, and $m$ is an auxiliary parameter denoting the number of inequivalent FE switching paths within one u.c. period. We use $\lambda$ rather than *x*, because the polarization change usually includes a complex symmetry-adapted mode changes, despite may be dominated by the displacing of one type of ion. Still, there are apparent analytical relations among expected <*P*>, <λ>, and <*x*> of the conducting ions. In this model, we assume that $\mathcal{F}(\lambda)$ is an even function, where λ = 0 state corresponds to the expected non-polar reference structure. This model includes the effect of second-order FE phase transition at $T_C$, above which there will only be direct ionic hopping among non-polar structures. In the following sections, we shall use this model for describing FE-HfO$_2$ as an example.

*Ion paths in FE-HfO$_2$.*—FE-HfO$_2$ has eight types of quasi- / pseudo-chiral unit cell (u.c.) variants [23,32,33]. Four variants are parallel and the other four are antiparallel polarized with the *c*-lattice, marked as S$_{1-4}$ and S$_{1'-4'}$, respectively. All variants can be mutually transformed via reflection / rotation operations [23,32] (e.g., $\hat{\sigma}_{ab} \times S_1 = S_{1'}$, $\hat{\sigma}_{bc} \times S_1 = S_2$, and $\hat{\sigma}_{ac} \times S_1 = S_4$), and the variants with same polarization directions can also be mutually transformed via translation operations [33] (e.g., approximately $(0.5, 0, 0.5) \times S_1 = S_2$, and $(0.5, 0.5, 0) \times S_1 = S_4$). Then, one can classify maximum 4 types of 180° FE switching paths by treating variants as the starting or ending point configurations, and 93 types of orthogonal domain walls (DW) [33]. These 4 paths are marked as {S$_1$↔S$_{1'-4'}$}, where we can always use the variant S$_1$ as starting point due to the symmetrical equivalence relations (e.g., $(0.5, 0, 0.5) \times \{S_2 \leftrightarrow S_{1'}\} = \{S_1 \leftrightarrow S_{2'}\}$). Note that such variant start-end paths alone do not define the atomic moving trajectories, until we further define the atomic start-end relations and perform atomic simulations. During the simultaneous switching in ideally clamped FE-HfO$_2$, three-coordinated active O-ions [34] can move in-between nearby Hf (002) planes, making the system traversing the *P*4$_2$/*nmc* tetragonal T-phase via {S$_1$↔S$_{2',4'}$}, or the $Fm\bar{3}m$ cubic C-phase via {S$_1$↔S$_{1',3'}$} [23,24,33,35]. Alternatively, they can move across Hf (002) planes, traversing the *Pbcm* phase via {S$_1$↔S$_{1'}$}, which is defined under the same variant start-end path as one of the above mentioned C-phase traversing path [23,35,36] (more backgrounds are in the supplementary material section one, SM-S1 [37]). These two sets of Hf (002) nonpenetrating / penetrating switching paths are head-to-tail connected, i.e., their atomic trajectories of active O-ions are connected and continuous [17,31]. They seem define the same structure as in oppositely polarized states with different high symmetry reference structures (*P*4$_2$/*nmc* T-phase and *Pbcm* phase).

To simplify the language, we mark those connected spatial trajectory paths in [001] as uniaxially-connected-paths (UCPs). And, we separately name these in FE-HfO$_2$ as UCP-a*I* (*I* = 1, 2, 3, 4) and UCP-b, which are Hf (002) nonpenetrating and penetrating paths, respectively.

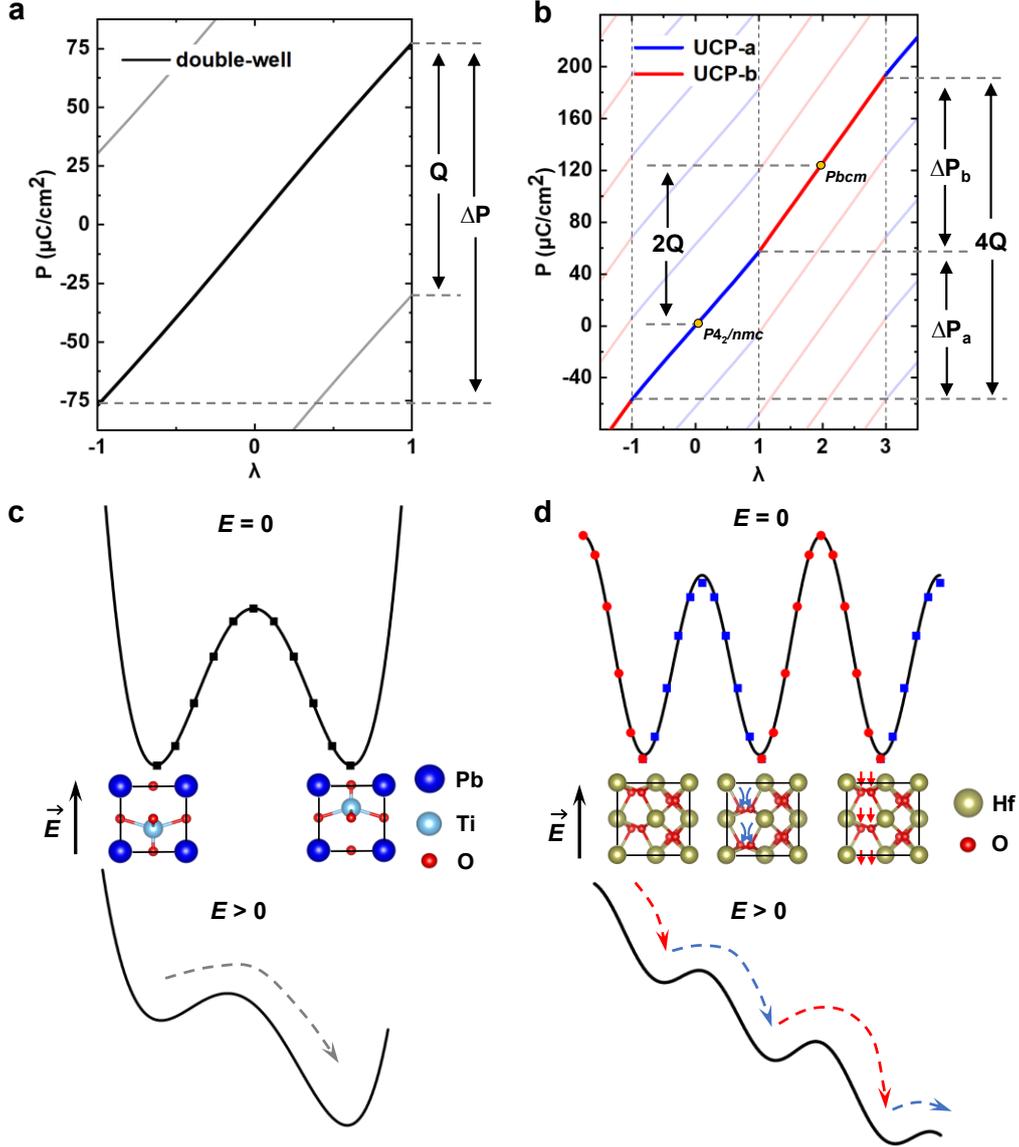

FIG. 1. Polarization and energy curves of double-well ferroelectrics and FE ionic conductor during simultaneous reversal. (a-b) MTP defined $P(\lambda)$ relations in PbTiO$_3$ and FE-HfO$_2$, respectively. Blue and red lines represent UCP-a2/a4 and UCP-b, respectively. Gold dots denote non-polar reference states. (c-d) the energy potential of PbTiO$_3$ and FE-HfO$_2$, respectively, under zero or finite electric-fields. The data points are calculated from density-functional-theory at zero field and Kelvin. The black solid curves are from the Landau-Devonshire expression or our proposed Eq. (2), with or without the inclusion of non-zero electric-fields. Insets indicate the periodic u.c. atomic structures at each energy minimum.

*Potential and Piezoelectricity of FE-HfO$_2$.*—Now we make a comparison of polarization and energy descriptions during ideal simultaneous FE switching, between double-well ferroelectrics PbTiO$_3$ and type-III FE ionic conductor FE-HfO$_2$. As plotted in Fig. 1a, FE-PbTiO$_3$ has infinite numbers of monotonic *P*-branches separated by *Q*. Due to the lack of a proper ion path, large polarization is generally considered leading to diverging energy (upper panel of Fig. 1c). With the energy unit meV and polarization unit $\mu C/cm^2$, Landau-Devonshire model [38] gives

$$\mathcal{F}(P, T = 0, E) = -0.035P^2 + 2.9 \times 10^{-6}P^4 - EP. \qquad (3)$$

Then, the application of a large enough $E$ can reverse $P$ by tilting the double-well (lower panel of Fig. 1c) and overcoming the barrier.

In FE-HfO$_2$, the UCPs permit long-range O-ion conduction. As shown in Fig. 1b, MTP still defines monotonic $P$-branches. The bold center-branch refers to the parent non-polar T-phase as the $P(\lambda) = P(0) = 0$ state. In the upper panel of Fig. 1d, its continuous comb-shaped energy profile is fitted by our model of Eq. (2) as

$$\mathcal{F}(\lambda, T = 0, E) = -0.12 \cos(\pi\lambda/2) + 0.59 \cos(\pi\lambda) - EP(\lambda). \qquad (4)$$

Considering the coupling $-EP$, a finite electric-field gives a stair-like potential enabling reversible ion migrations in fundamental (lower panel of Fig. 1d).

We confirm the "spontaneous polarization" [23,36] of UCP-a is $P_a = \Delta P_a/2 = 0.57\ C/m^2$, and that of UCP-b is $P_b = \Delta P_b/2 = 0.68\ C/m^2$. Between two nearest centrosymmetric states ($P4_2/nmc$ T-phase and $Pbcm$ phase), we find $\Delta P = 2Q$, since there are four O-ions being transported for 1/4 lattice constants. This indicates the -2 oxidation state of O-ions in HfO$_2$ [23,39]. Consistently, $\Delta P_a + \Delta P_b = 4Q$, where four O-ions move for 1/2 lattice constants and $\rho(\mathbf{r})$ is left unchanged.

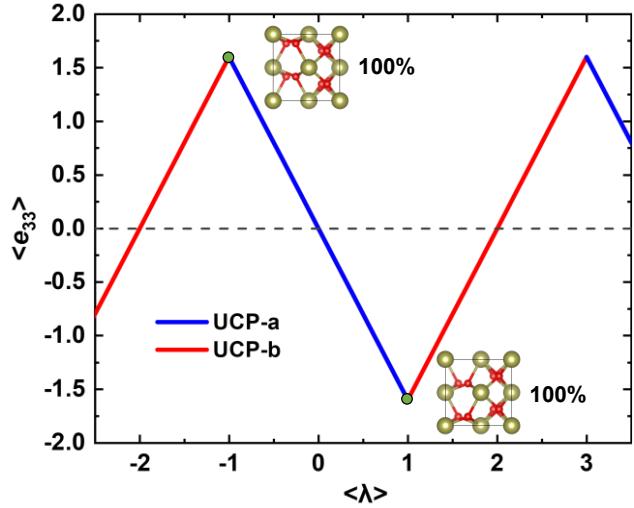

FIG. 2. Relation between the expected piezoelectric coefficient $<e_{33}>$ and polarization switching coordinate $<\lambda>$ in successive FE switching. The up-polarized state defined in UCP-a path show negative total piezoelectric coefficient $e_{33}$ = -1.54 C/m$^2$, close to previous reports [40-42].

On the other hand, the piezoelectric response should better be considered in an asynchrony switching scheme [16]: unit cells exhibit different piezoelectric responses when they are in different polarization states and electric-field directions, and the stochastic average of their contributions give the measured piezoelectric coefficient of the detected area. During successive FE switching, the same sample may exhibit oscillating polarization-strain coefficient [16] $e_{33}$, originating from the phase-independent nature of proper piezoelectric response [43]. We propose that the situation in HfO$_2$ should be similar, as plotted in Fig. 2: piecewise lines indicate opposite local piezoelectric phase loops in HfO$_2$, depending on the switching path.

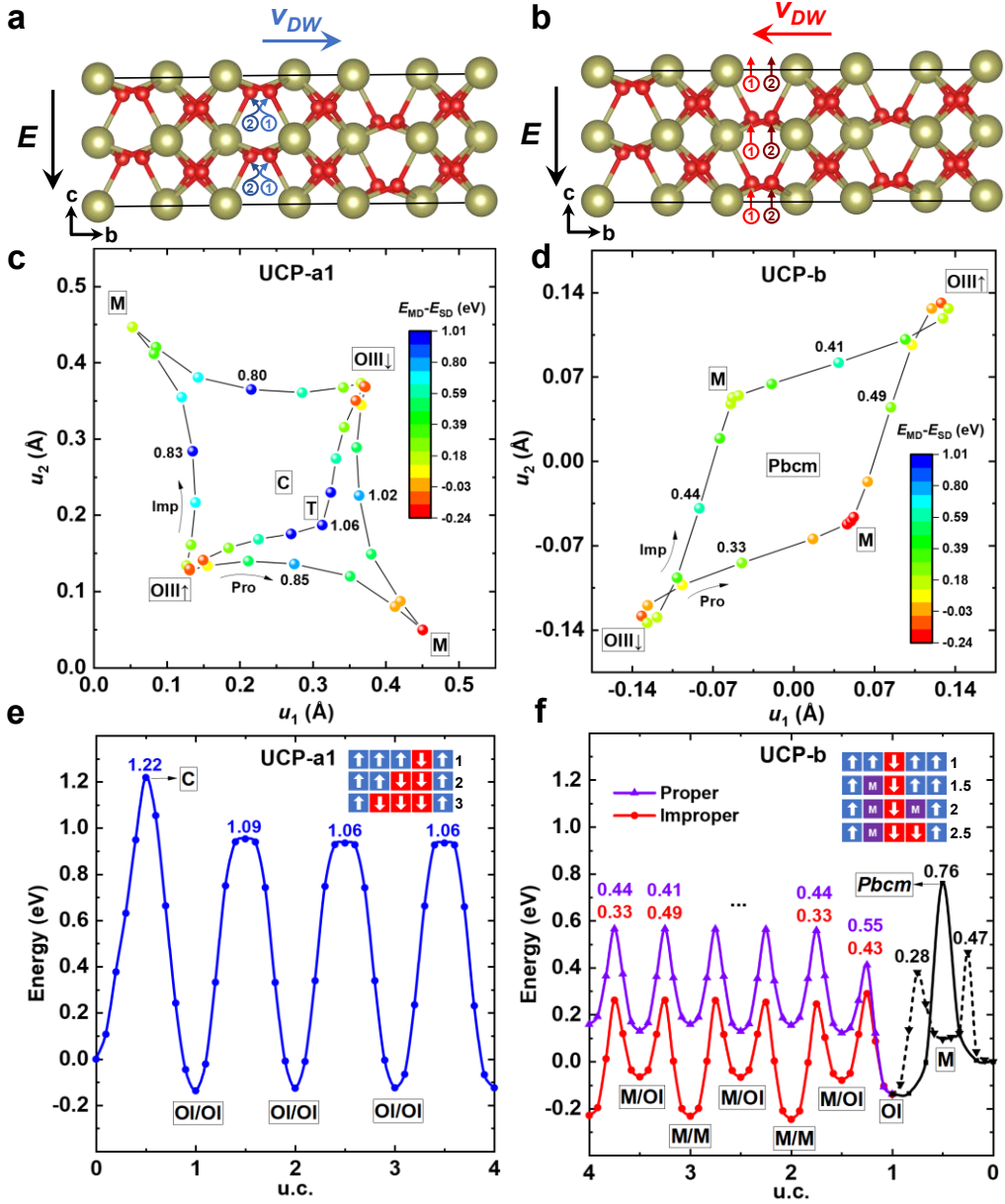

FIG. 3. Opposite motions of shoulder-to-shoulder 180° DW via either UCP-a1 or UCP-b along [010]. (a-b) DW motions via UCP-a1 and UCP-b under $E \parallel -z$, respectively. The O-ion rows labeled "1" and "2" are found moving asynchronously. (c-d) zero-field, zero-temperature MEPs of DW motion in a $1 \times 8 \times 1$ multidomain supercell via UCP-a1 and UCP-b, respectively. The coordinate $u$ corresponds to the O-ion $z$-displacement from their nearest Hf (002) planes. The total energy of a single domain structure is taken as the zero reference. Elementary barriers are labeled in unit eV. The paths where row-1 move with priority are marked as proper paths, while those with row-2 taking privileges as improper paths. (e-f) nucleation-and-growth MEPs via UCP-a1 (properly through the T-phase) and UCP-b (properly or improperly through the M-phase), respectively. Local phases of typical points are labeled. Insets indicate corresponding u.c. configurations.

*Collective ionic conduction via nucleation-and-growth.*—It is known that a displacive FE switching generally happens in the form of nucleation-and-growth [44]. New domains nucleate, grows quickly vertically and then creepingly horizontally, and finally merge to complete a full FE switching. Here, we propose a physical model of ionic conduction via nucleation-and-growth in displacive FE-HfO$_2$. This may be considered an FE extension to the domain model of van Gool [45,46] for explaining anomalously fast diffusion.

There have been relevant theoretical researches performed under a simpler picture in thin film FE-HfO$_2$ [23,47]: all vertical nucleation-and-growths happen at once in very thin films (concerted switching) with the smallest nucleus size, and domains grow horizontally via DW motion in whole rather than in steps [48,49]. Our first-principles calculations would be under the same assumption to present a basic intuitive picture. UCP-a1 has been considered important, since it may lead to a favored shoulder-to-shoulder type DW with weakly negative DW energy [22,23,33,47] ($E_{DW}$ = -2.4 meV/Å$^2$ in our calculations). This often-reported negative $E_{DW}$ originates from the thermodynamically unstable nature of the FE OIII-phase, which is considered rather kinetically stabilized [22,50]. The local DW structure is generally considered as in OI-phase (*Pbca*) [47], whose u.c. consists of two oppositely-polarized OIII-u.c. (OI = OIII ↑ +OIII ↓). A sluggish DW motion mechanism along [010] was proposed based on this DW formed via UCP-a1 [47], while much faster DW motion mechanisms along [010] were proposed based on DWs formed via UCP-a2 or UCP-a4 [23,33], albeit with positive $E_{DW}$. And, Du et al. [51,52] argued that the local structure of shoulder-by-shoulder DW formed via UCP-a1 should be the monoclinic M-phase (*P2$_1$/c*) with even lower $E_{DW}$ than that of OI-phase, evidenced by first-principles calculations and experimental comparisons. If we stack these M-phase DW structures, and we would get the second type of *Pbca* phase (OV) [22,53]. UCP-b and UCP-a1 are both {S$_1$↔S$_{1'}$} in the sense of u.c. variant start-end path. Meanwhile, they are conflicting in determining the direction DW motions (see opposite DW propagation vectors in Fig. 3a-3b).

With first-principles calculations, we investigate the elementary nucleation-and-growth process in a fully polarized bulk FE-HfO$_2$, by calculating zero-field minimum energy paths (MEPs). The energy barriers $\Delta E$ and DW energies $E_{DW}$ are directly related to the activation energy and critical field of FE switching [23,48,49,54,55]. Interestingly, the exchange of moving privileges between two active O-ions rows (displacements $u_1 \neq u_2$) in the DW (Fig. 3a-3b) introduces two different proper and improper kind of paths (defined in Fig. 3c-3d). Improper paths have overall higher total energies than that of proper paths, but the growth barriers $\Delta E_{grow}$ of improper paths can be lower than that of proper ones.

For DW motions via UCP-a1 (Fig. 3a), imbalanced Hf (002) planes on two sides of DW lead to a T-phase intermediate structure [56] (two inner paths in Fig. 3c), rather than the C-phase in fixed lattice simulations [24,33]. As shown in Fig. 3e, growth barrier $\Delta E_{grow}$ = 47 meV/Å$^2$ is lower than the nucleation barrier $\Delta E_{nuc}$ = 41 meV/Å$^2$. Traversing M-phase via UCP-a1 (two outer paths in Fig. 3c) may bring lower $E_{DW}$ = -4.4 meV/Å$^2$ and $\Delta E_{grow}$ = 32~33 meV/Å$^2$. However, this is unlikely to happen, since a O-ions row must move against the driving force of $E$ (negative slope in the $u_1$-$u_2$ curves).

Oppositely, UCP-b (Fig. 3f) is found energetically preferred with a lower $\Delta E_{nuc}$ = 30 meV/Å$^2$ through the *Pbcm* phase, or 18 meV/Å$^2$ traversing the M-phase. Then during growth (Fig. 3d), we find the M-phase DW emerges, and has a surprisingly low rate-determining motion barrier $\Delta E_{grow}$ = 19 or 17 meV/Å$^2$ in in proper and improper configurations, respectively. Both $\Delta E_{nuc}$ and $\Delta E_{grow}$ of UCP-b are lower than that of UCP-a1 in [010]. If we take T$_C$ = 600 K, the corresponding nucleation and growth critical electric-fields via UCP-b are estimated according to $E_c(0) \approx \Delta E/\Delta PV_{u.c.}$ (where V$_{u.c.}$ is the volume of u.c.) and $E_C(T) = [(T_C - T)/T_C]^{3/2} \times E_c(0)$ [23] as 5~6 MV/cm at room-temperature, significant lower than 8~10 MV/cm via UCP-a1. They are also expected to be lowered by doping, defect, and interface effects. Extending investigations of DW motions along [010] ([100]) are referred to SM-S2 (SM-S3) [37].

Overall, we find UCP-b disables UCP-a1 in DW motions of stoichiometric FE-HfO$_2$, changing the prevailing picture of sluggish DW motion via UCP-a1 [47]. The existence of new ground M-phase DW is also validated to be energetically favored over the OI-phase. It is possible to experimentally confirm the O-ion motions via UCP-b, if any *E*-driven motion of a M-phase DW is observed. Other UCP-a*I* paths (SM-S4 [37]), except for UCP-a1, would be responsible for conductions within nearby Hf (002) planes.

An intuitive picture of successive FE switching in displacive FE-HfO$_2$ would be like Fig. 4a: under an external electric-field *E*, new domains are infinitely emerging and growing from the inside of other oppositely-polarized domains. *Such a domain nesting cycle shall repeat until E is canceled out*. In this way O-ions are conducted collectively. Vacancies or excesses of active ions may further stabilize the DW and lower their migration barriers [12,57]. We think our picture may provide an alternative explanation to observations of mixed domain pattern with opposite piezoelectric phase loops [58].

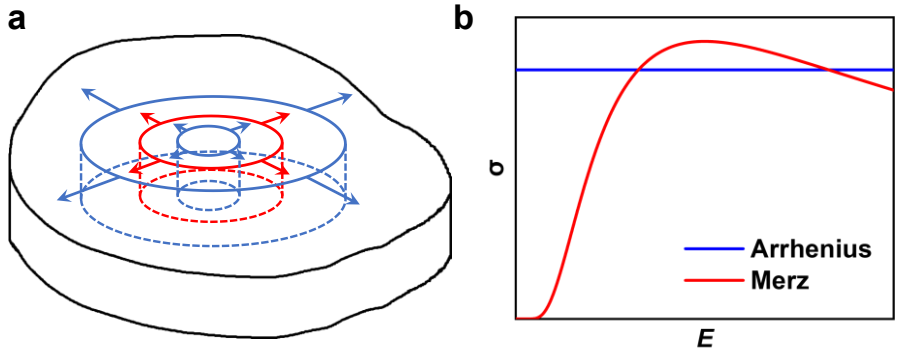

FIG. 4. (a) A sketch showing the nesting-domain pattern in FE-HfO$_2$ during successive FE switching. (b) The difference on the field-dependency between Merz and pristine Arrhenius type ionic conductivities. The relative magnitudes are meaningless.

*Ionic conductivity.*—Interfaced with ion-reservoirs as source and drain, FE ionic conductors like HfO$_2$ may act as electrolytes with significant features. Ion migration is supposed to be achieved synergically via the nucleation-and-growth procedure, and such a collective feature is consistent with that of super-ionic conductors [59]. The field-dependency of ionic conductivity $\sigma$ (Fig. 4b) can also be nontrivial: a generally taken form is in the pristine field-unrelated Arrhenius type, which can be estimated using the Nernst-Einstein relation with individual ion migration deduced diffusion coefficient [60]:

$$\sigma_{Arrh} \sim T^{-1} e^{-\Delta U/T}, \tag{5}$$

where $\Delta U$ is the activation energy, and $T$ the temperature. Other parameters are omitted for simplicity. On the other hand, $\sigma$ of FE ionic conductor is supposed to obey the Merz's law [49,54] describing FE switching rates (extended discussion see SM-S5 [37]):

$$\sigma_{Merz} \sim E^{-1} e^{-\Delta U/ET}. \tag{6}$$

Such a difference is attributed to that FE switching has strong ionic correlations, which are usually omitted or underestimated. Similiar nonlinear $\sigma - E$ relations were reported in the studies of strongly correlated ionic conductivity [61,62], referring to the Onsager-Wien effect. It is interesting to find that the simple mathematical form of Merz's law fits well in describing the nonlinear effect of ionic conductivity, similar to the Mott-Gurney law pplied in ionic conduction [63]. Importantly, such a conductivity suggests a good robustness to perturbations in artificial synapses.

*Final remarks.*—To conclude, we present a free-energy model for FE ionic conductors that may be implemented to multiscale simulations after further developments. We propose a FE-extended domain model for describing the ionic conduction in displacive FE ionic conductors (e.g., FE-HfO$_2$), and further infer an associate nesting domain pattern with mixed piezoelectric responses during FE switching. The associate ionic conductivity should be field-dependent following the Merz's law. The effects of defect, film thickness, and grain size await further investigations, since they should be important for the realistic switching dynamics and performances in FE ionic conductor based devices. It will also be interesting to investigate the ionic conductivity of FE-HfO$_2$ and other similar FE ionic conductor [36], in pursue of low temperature fast oxide-ion conductors [59,64] and artificial synapses.

*Methods.*—Density-functional-theory calculations are based on the projected augmented wave method implemented in the Vienna ab initio Simulation Package (VASP) [65,66] with a plane-wave-basis cutoff of 600 eV. The used exchange-correlation functional is local-density-approximation [67]. We included 10 valence electrons for Hf, and 6 for O element in pseudopotentials. Spacing in Γ-centered Monkhorst-pack *k*-grids is less than 0.3 Å$^{-1}$. All constructed DWs consist of $1 \times 8 \times 1$ u.c. building blocks (a = 5.162, b = 4.960, and c = 4.978 Å), being fixed in lattice constants during relative atomic coordinate relaxing until Hellmann–Feynman forces are below 5 meV Å$^{-1}$. $E_{DW} = (E_{MD} - E_{SD})/2S$, where $E_{MD}$ and $E_{SD}$ are multi-domain and single-domain energies, respectively, and $S$ denotes the DW area. Polarization values are evaluated with the Berry-phase method [68]. The DW motion barriers are calculated via the climbing-image nudged-elastic-band (CI-NEB) method[69]. Atomic plots are from the VESTA software [70].

X.Z. thanks the Shanghai Sailing Program (20YF1401700). X.L. gives special thanks to the Research Start-up Fund Project of Shaoxing University and Natural Science Foundation of Zhejiang Province (LQ23A040003). W.R. thanks the support by the National Natural Science Foundation of China (12074241, 11929401, 52130204), the Science and Technology Commission of Shanghai Municipality (19010500500, 20501130600), High Performance Computing Center, Shanghai University, and Key Research Project of Zhejiang Lab (2021PE0AC02).

*xiaona_zhu@fudan.edu.cn

# Supplemental Materials of
# Polarization description of successive ferroelectric switching in hafnia

Guo-Dong Zhao,[1] Xingen Liu,[2] Zhongshan Xu,[1] Wei Ren,[3] Xiaona Zhu,[1,4]* David Wei Zhang,[1,5] and Shaofeng Yu.[1,5]

[1]*School of Microelectronics, Fudan University, Shanghai 200433, China*
[2]*School of Mathematical Information, Shaoxing University, Shaoxing 312000, China*
[3]*Physics Department, Shanghai Key Laboratory of High Temperature Superconductors, State Key Laboratory of Advanced Special Steel, International Centre of Quantum and Molecular Structures, Shanghai University, Shanghai 200444, China*
[4]*Jiashan Fudan Institute, Jiashan 314100, China*
[5]*National Integrated Circuit Innovation Center, Shanghai 201204, China*

## Outline

## Section S1. Elementary FE switching paths in $HfO_2$

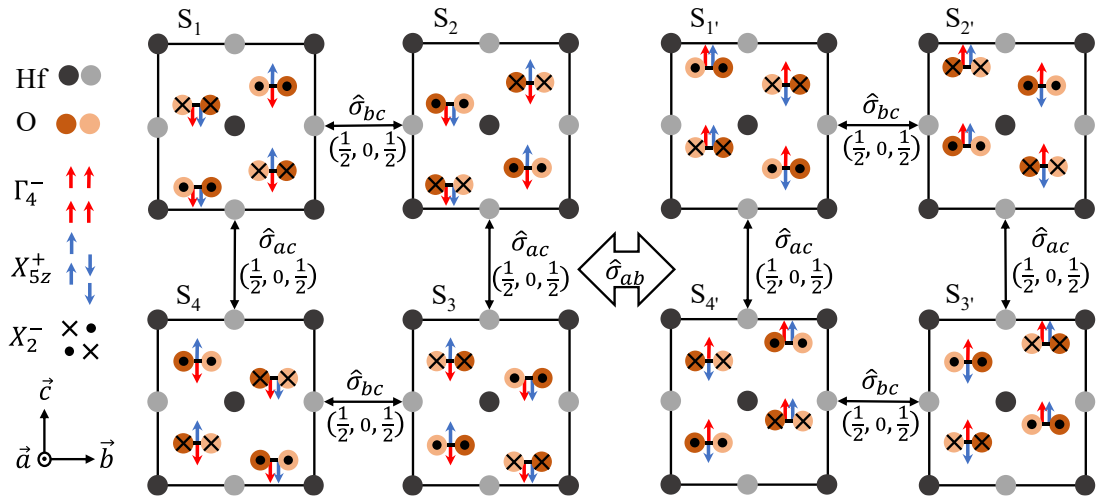

Fig. S1. A sketch of eight energy degenerate OIII-$HfO_2$ unit cells with upward ($S_{1-4}$) or downward ($S_{1-4}$) polarization in the definition of UCP-a paths. Their mutual transformation relations are labeled for nearby variant cells. The grey (orange) spheres represent Hf (O) atoms, with darker (lighter) color denoting their closer (farther) distance to the eyes. Red, blue, and black (● and ×) arrows represent the symmetry-adapted lattice distortion modes of $\Gamma_4^-$, $X_{5z}^+$, and $X_2^-$, respectively. The distortions of Hf atoms are not shown in the sketch. Other mode components $X_{5y}^+$, $X_{5z}^+$, $X_5^-$, and $X_3^-$ are not labeled in the sketch.

As concluded in our previous work [1], there are 8 unit cell variants of $Pca2_1$-$HfO_2$ if we omit its translation symmetry (Fig. S1). Then, we can find 4 inequivalent FE switching paths, $\{S_1 \leftrightarrow S_{1'\text{-}4'}\}$, in the term of initial and final points. The switching paths, considering only the oxygen ions moving within their own octants, were already studied. Oppositely, if we consider the oxygen ions moving across the Hf(002) planes, there are in principle another 4 possible paths.

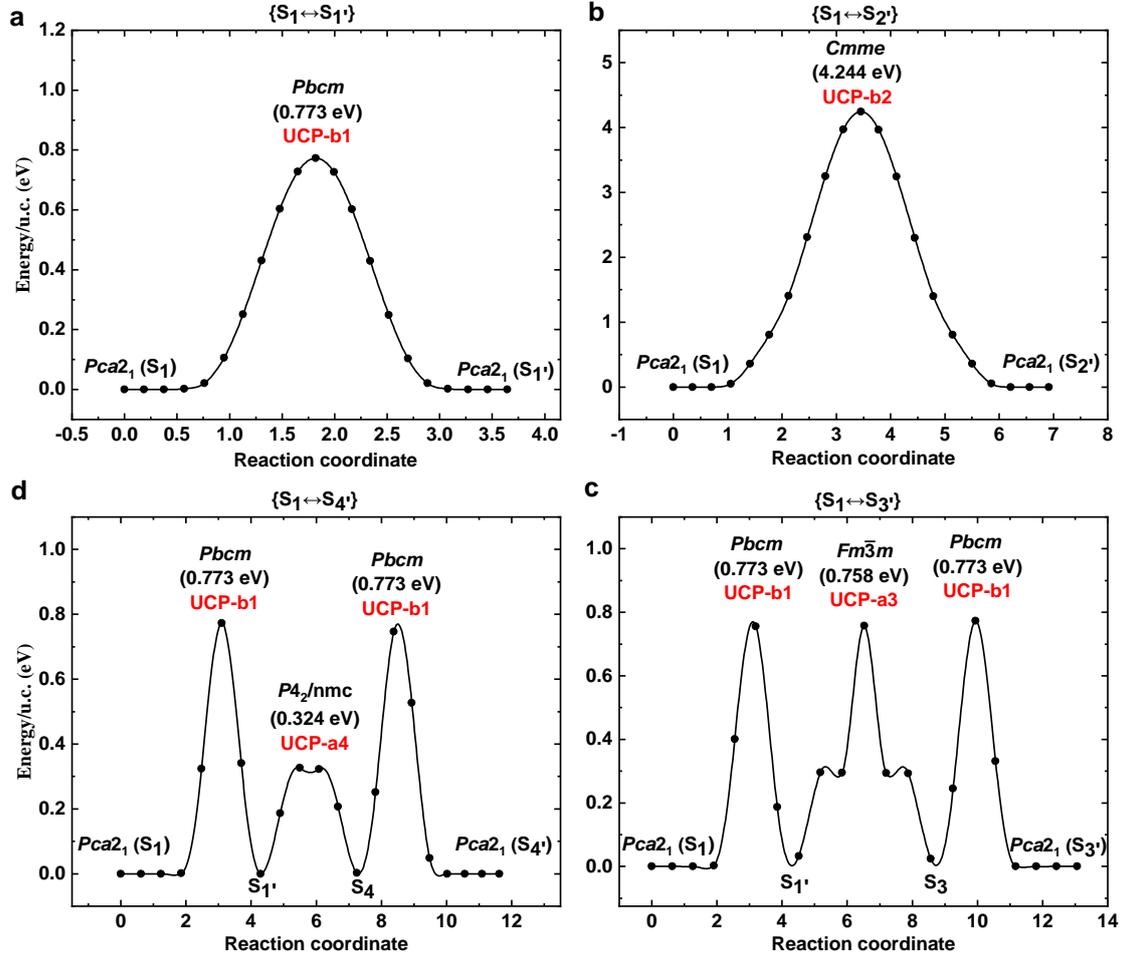

Fig. S2. Homogeneous FE switching penetrating Hf(002) planes. (a-d) the MEPs of $\{S_1 \rightarrow S_{1'\text{-}4'}\}$ paths penetrating Hf(002) planes, respectively. Space groups of specific transition points are labeled. The highest energy barriers of $\{S_1 \rightarrow S_{1'\text{-}4'}\}$ paths are 0.77, 4.23, 0.77, and 0.77 eV, respectively.

As shown in Fig. S2, the MEPs of $\{S_1 \leftrightarrow S_{1'\text{-}4'}\}$ paths penetrating Hf(002) planes are calculated with the CI-NEB method [2] in a u.c. with fixed lattices. The first path in Fig. S2a, $\{S_1 \leftrightarrow S_{1'}\}$ traversing a structure with $Pbcm$ space group symmetry, is denoted as UCP-b in our main-text, and here UCP-b1, specifically. The second path shown in Fig. S2b, $\{S_1 \leftrightarrow S_{2'}\}$ traversing a high energy $Cmme$ structure, is denoted as UCP-b2 in this supplementary material. The atomic details of UCP-b1 and UCP-b2 will be illustrated below (Fig. S3-S4). The other two paths in Fig. S2c-S2d, $\{S_1 \leftrightarrow S_{3',4'}\}$, are meaningless since they turn out to be just combinations of UCP-b1 and UCP-a paths.

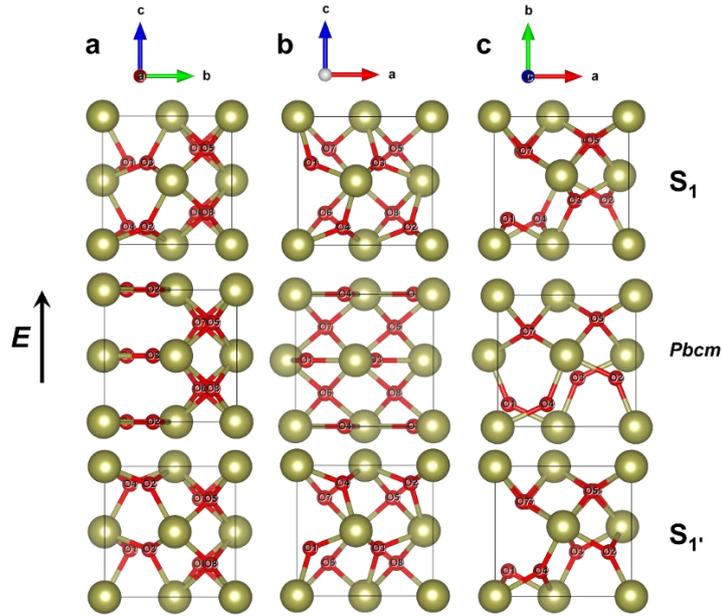

Fig. S3. The initial, saddle, and final point structures of UCP-b2 in the homogeneous FE switching scheme. (a-c) views along *a*, *b*, and *c*-axes, respectively.

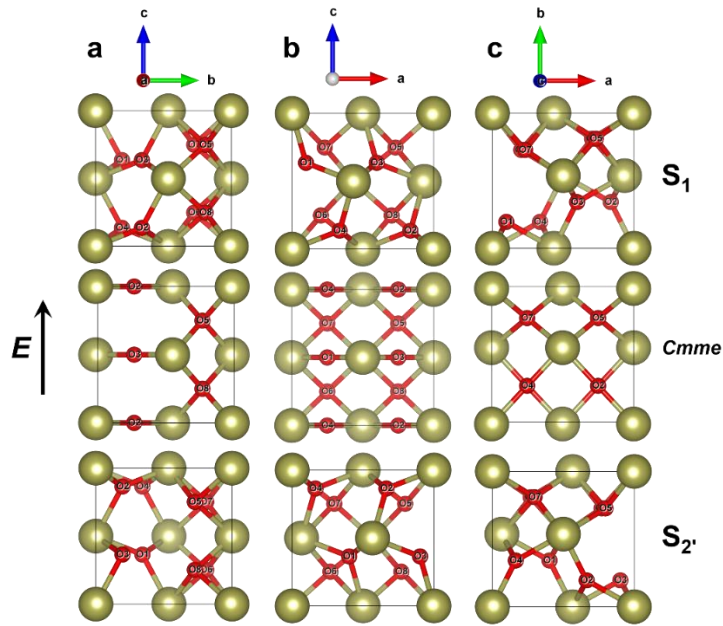

Fig. S4. The initial, saddle, and final point structures of UCP-b2 in the homogeneous FE switching scheme. (a-c) views along *a*, *b*, and *c*-axes, respectively.

Via the UCP-b1 path, active oxygen ions penetrate Hf(002) planes symmetrically (for example, O1 and O3 moves synchronously). While it is also possible that move asymmetrically due to random thermal fluctuations or asymmetric defects. More discussions are in Sec. S2. Via the UCP-b2 path, the cell must travers a high symmetry, high energy *Cmme* transition structure. Its barrier may be lowered by taking the path as $\{S_1 \rightarrow S_{1'} \rightarrow S_{2'}\}$, but this is impossible since the unidirectional external electric-field cannot reverse the nonpolar modes $X_2^-$, $X_5^-$, and $X_{5y,z}^+$, which are opposite in $S_1$ and $S_{2'}$ cells. To conclude, only the UCP-b1 worth considering.

## Section S2. UCP-b1 along [010]: the 180° shoulder-to-shoulder SS$^+$1' DWs

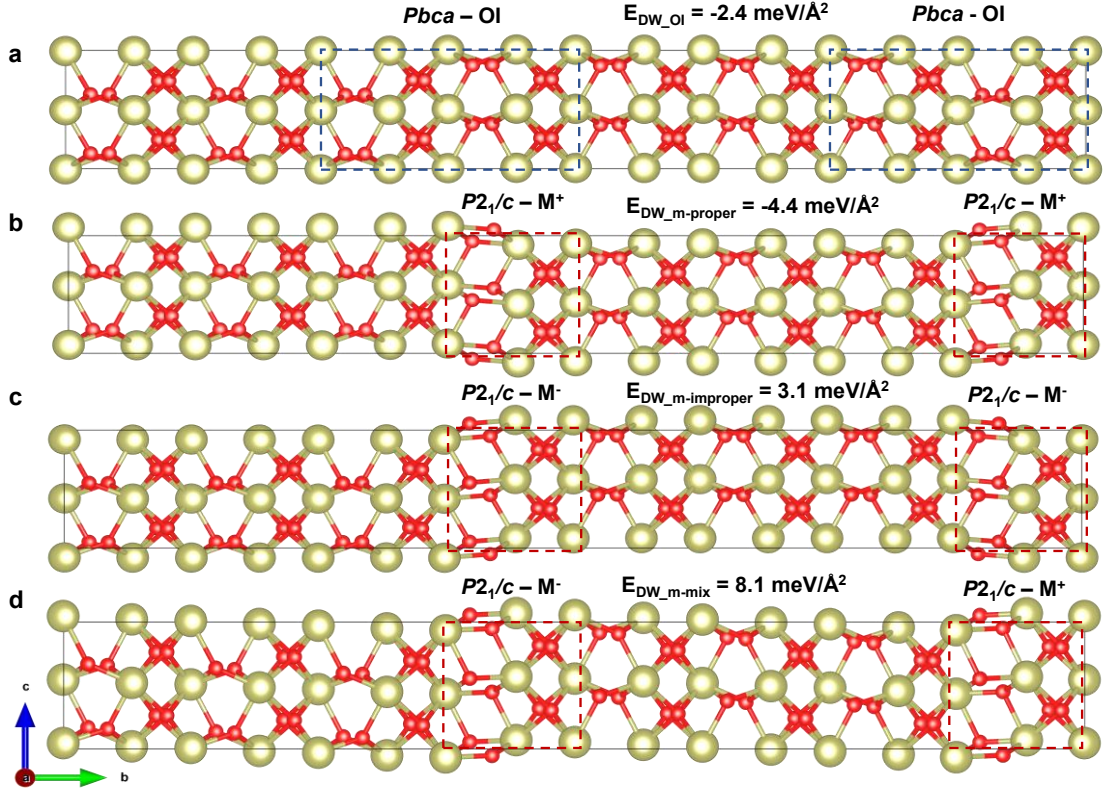

Fig. S5. The atomic structures of evenly spaced (a) *Pbca* OI-phase and (b) proper, (c) improper, and (d) mixed *P2$_1$/c* M-phase 180° shoulder-to-shoulder SS$^+$1' DWs in a $1 \times 8 \times 1$ periodic supercell. The corresponding DW energies E$_{DW}$ are labeled.

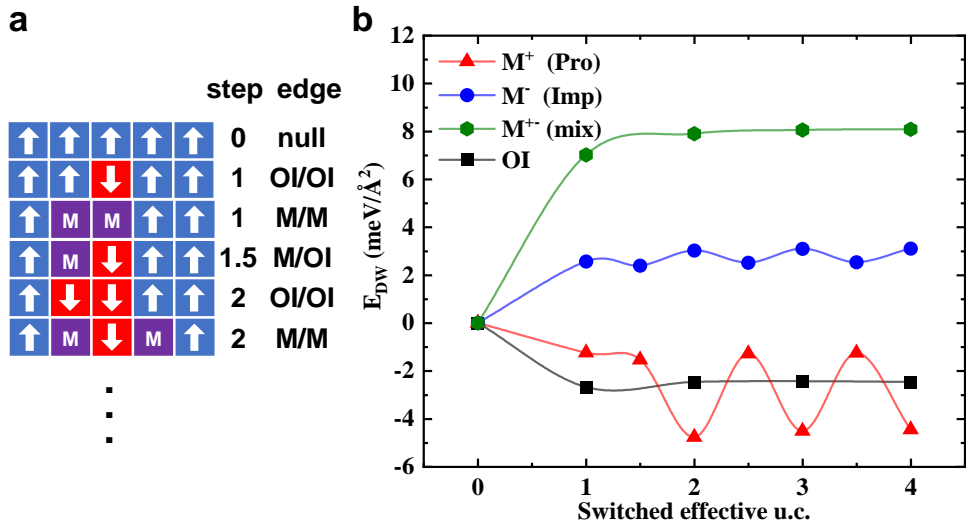

Fig. S6. (a) DW configurations of OI and M-phase (generated by proper, improper, and mixed switching UCP-b1 paths) DW structures and (b) their DW energies E$_{DW}$, calculated by step-wisely switching unit cells within a $1 \times 8 \times 1$ periodic supercell. Details of the initial steps are shown in Fig. S7a. The fitting curves are meaningless.

All 93 types of orthogonal DWs of orthorhombic $HfO_2$ had been comprehensively classified with symbolic notations in our proposed general DW definition [1]. We focus on the low strain 180° side DWs (neutral), which is dominant in FE thin films [3].

In this work, we investigated as many as possible SS$^+$1' 180° DW atomic structures in their stabilities and emergences via UCP-a1 and UCP-b1 paths. Du et al. [4] proposed monoclinic local DW structure is highly possible based on their low domain wall energies $E_{DW}$ and consistencies with transmission electron microscopy images. As discussed in our main-text, monoclinic DW structure may emerge during motions of SS$^+$1' DW via the UCP-b1 path. However, the detailed moving order of active oxygen ion rows leads to different monoclinic local DW structures and different $E_{DW}$.

As mapped in Fig. S5, the proper configuration (Fig. S5b) of $P2_1/c$ M-phase DW structure indeed leads to a lower $E_{DW}$ than that of $Pbca$ OI-phase, while the improper and mixed configurations (Fig. S5c-d) both in turn lead to higher and positive $E_{DW}$.

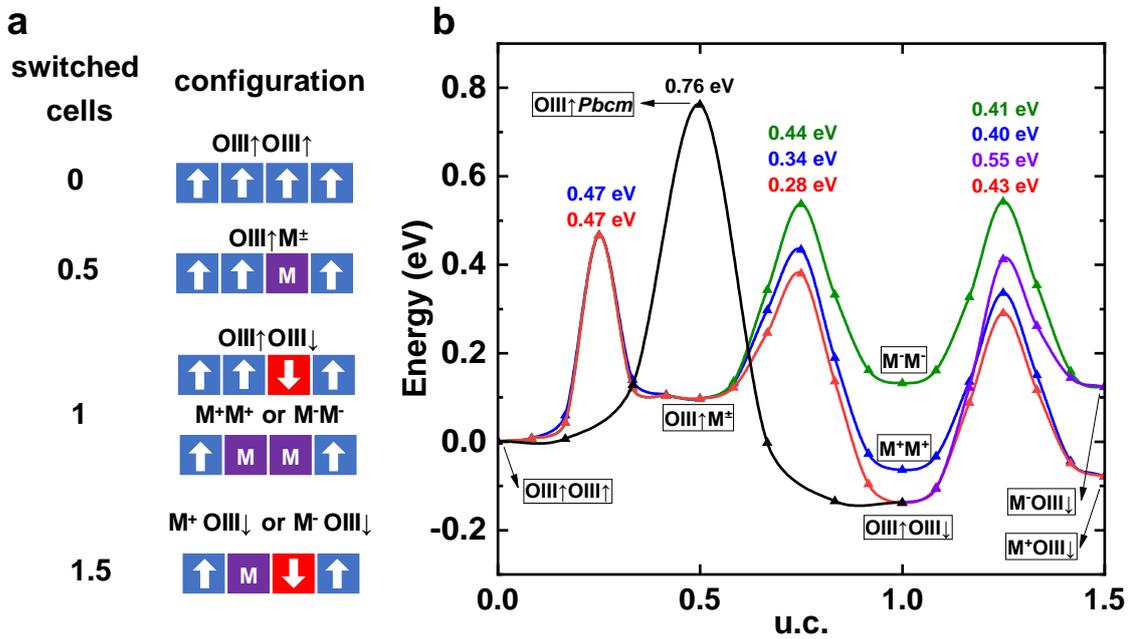

Fig. S7. Cell-by-cell nucleation in four possible ways via UCP-b1 along [010], calculated in a $1 \times 8 \times 1$ periodic supercell. (a) configuration sketch of 4 in 8 cells; (b) the corresponding MEPs. The symmetries of 2 specific cells that will be first switched in the model are labeled.

The initial MEP shown in the main-text Fig. 3(f) (effectively switching from 0 to 1.5 cells) is symmetric where two rows of active oxygen ions (labeled in the Fig. 2a-2b of main-text) move synchronously with balance hafnium planes on both sides. But considering the random thermal fluctuations or asymmetric defects, asynchronous moving of oxygen rows is allowed (Fig. S7). This would lead to much lower barrier (0.47 eV) than that traversing a $Pbcm$ structure (0.76 eV). This means that the event probability of an "opposite" domain nucleation during switching can be high. Based on the concept in generally applied nucleation limited switching (NLS) model, that the it is the rate of comparably slow nucleation limiting the total switching rate, the record low nucleation barrier of UCP-b1 is expected to make UCP-b1 more general than UCP-a [5] in experiment observations.

We find the detail MEPs are intricate (see also in Fig. S7) in the specific comparisons among proper, improper, and mixed configurations. The initial switching from 0 to 0.5 effective cells is degenerate between proper and improper configurations (0.47 eV). Then from 0.5 to 1 effective cell switching: totally switch the M-cell into an oppositely polarized OIII-cell leads to the lowest barrier (0.28 eV) and lowest total energy; if a nearby cell is transformed into another M-cell, it would lead to a higher barrier (properly 0.34 eV, and improperly 0.44 eV) and higher total energy. During the switching from 1 to 1.5 effective cells: turning $M^{\pm}M^{\pm}$ into $M^{\pm}OIII\downarrow$ configurations lead to comparably lower barriers (properly 0.40 eV, and improperly 0.41 eV); turning $OIII\uparrow OIII\downarrow$ into $M^{\pm}OIII\downarrow$ configurations lead to comparably higher barriers (properly 0.43 eV, and improperly 0.55 eV).

Combining Fig. S6, S7 and Fig. 2f, the proper path would be dominant due to its lower initial nucleation barriers. The advantage in domain wall energy of proper path is also more obvious than its disadvantage in motion barrier. On the other hand, if the initial nucleation mechanism described in Fig. S7 is affected by interfaces, defects, or other factors, we would not be able to, based on the data we have, judge whether proper or improper paths (Fig. 2d and 2f) is dominant. To solve this issue, relevant high resolution microscopic imaging, reliable force field, and further first-principles investigations are required.

### Section S3. UCP-b1 along [100]: the 180° face-to-face $FF^{+}1'$ DWs

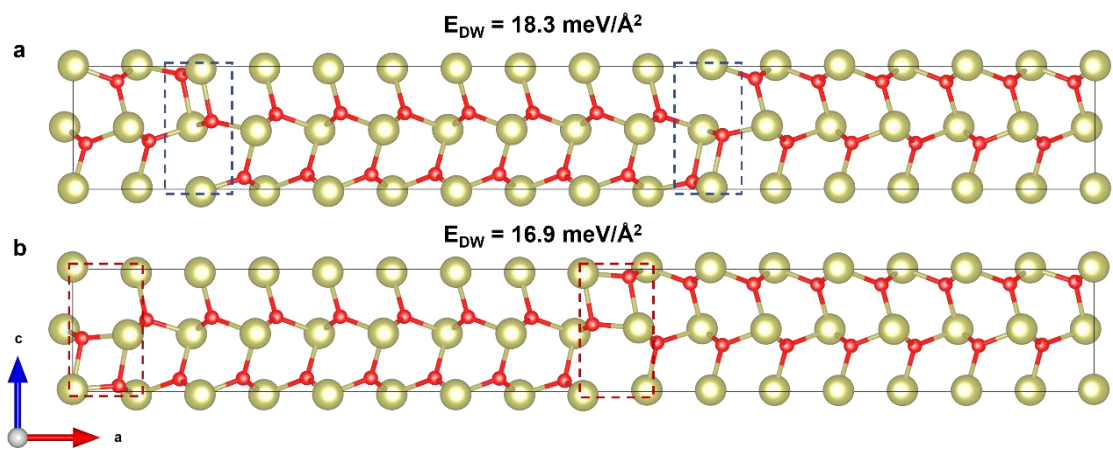

Fig. S8. Two possible 180° $FF^{+}1'$ DW ground structures (a) $G_1(FF^{+}1')$ and (b) $G_2(FF^{+}1')$ constructed in a $1 \times 8 \times 1$ periodic supercell, together with their corresponding $E_{DW}$. Only the three coordinate oxygen ions are shown for the convenience of eyes.

We find a new face-to-face $FF^{+}1'$ DW ground structure $G_2(FF^{+}1')$ (Fig. S8b), lower in energy than that of our previous studied one [1] (Fig. S8a). Its discovery in this work is due to the fact that it can only emerge during the DW motion calculations via UCP-b1. Only via UCP-b1, two active oxygen ions can share the same u.c. "octant" as framed in Fig. 8b, after penetrating the Hf(002) planes.

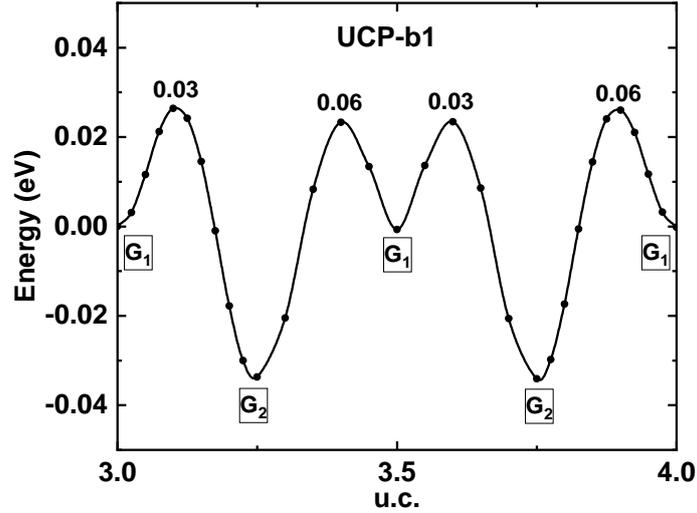

Fig. S9. The MEP of $G_1(FF^+1')$ DW motion via UCP-b1 along [100] in a $1 \times 8 \times 1$ periodic supercell. Clearly, it will inevitably fall partially back into $G_2(FF^+1')$. The motion barrier of each elementary step is labeled.

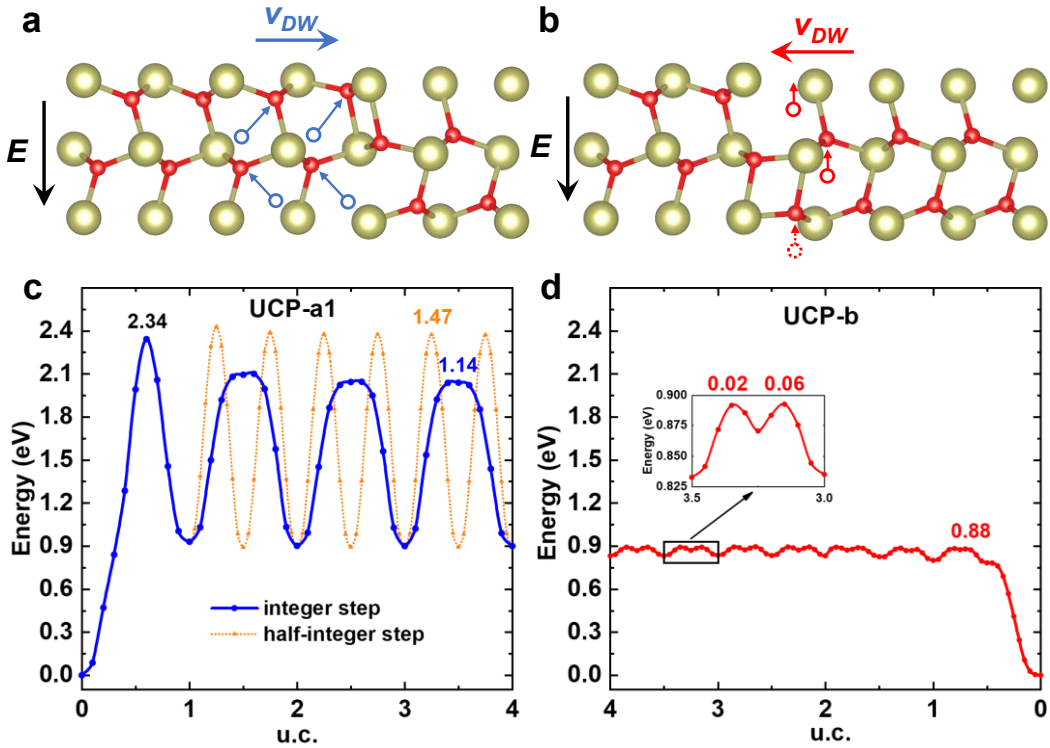

Fig. S10. Opposite motions of $FF^+1'$ DW within different UCPs along [100]. (a-b) motions of $FF^+1'$ via UCP-a1 and UCP-b under $\boldsymbol{E} \parallel -\boldsymbol{z}$, respectively. (c-d) nucleation-and-growth MEPs via UCP-a1 (with integer or half-integer u.c. steps) and UCP-b (with half-integer u.c. steps), respectively. Only the three coordinated O-ions are shown for the convenience of eyes.

In fact, we think it is also necessary to investigate the competence between UCP-a1 and UCP-b along the [100] direction, due to the anisotropic nature of $HfO_2$ [1]. As illustrated in Fig. S10a-S10b, UCP-a1 and UCP-b1 also result in opposite DW motions. And, UCP-b1 leads the new ground FF$^+$1' DW structure with slightly lower $E_{DW}$ (16.9 meV/Å$^2$) than that led by UCP-a1 (18.3 meV/Å$^2$). It is also evidenced in Fig. S10c-S10d that both $\Delta E_{nuc}$ (0.88 eV) and $\Delta E_{grow}$ (0.06 eV) of UCP-b are lower than that of UCP-a1 (2.34 and 1.14 eV, respectively) in [100].

**Section S4. UCP-a along [100] and [010]: other 180° SS$^+$ and FF$^+$ side DWs**

Table S1. DW energies $E_{DW}$ and motion barriers $\Delta E_{grow}$ (in $1 \times 8 \times 1$ supercell) of UCP paths (each corresponds to 2 orthogonal DW configurations) are calculated with either free or fixed lattice constants. Note that underlined values are calculated with pure OI-phase DW structures, while traversing M-phase intermediate structures.

| Ini-Fin Path | Atomic Path | DW | $E_{DW}$ (meV/Å$^2$) | | $\Delta E_{grow}$ (meV) | |
|---|---|---|---|---|---|---|
| | | | Free[1] | Fix | Free[1] | Fix |
| {S$_1$↔S$_{1'}$} | UCP-a1 | SS$^+$1' | -2 | **-2** | <u>972</u> | **<u>970</u>** |
| | | FF$^+$1' | 18 | **18** | 1148 | **1142** |
| {S$_1$↔S$_{2'}$} | UCP-a2 | SS$^+$2' | 28 | **29** | 79 | **81** |
| | | FF$^+$2' | 16 | **17** | 219 | **222** |
| {S$_1$↔S$_{3'}$} | UCP-a3 | SS$^+$3' | 24 | **25** | 693 | **750** |
| | | FF$^+$3' | 50 | **53** | 186 | **230** |
| {S$_1$↔S$_{4'}$} | UCP-a4 | SS$^+$4' | 25 | **26** | 199 | **214** |
| | | FF$^+$4' | 32 | **33** | 23 | **25** |

In this work, our first-principles density functional theory calculations are all carried out with fixed lattice constants. This is to approach the physics of thin film systems clamped by substrates, with reasonable computing resource consumption. As listed in Table S1., results obtained from such calculations are close and consistent with those with free lattice constants in our previous work [1].

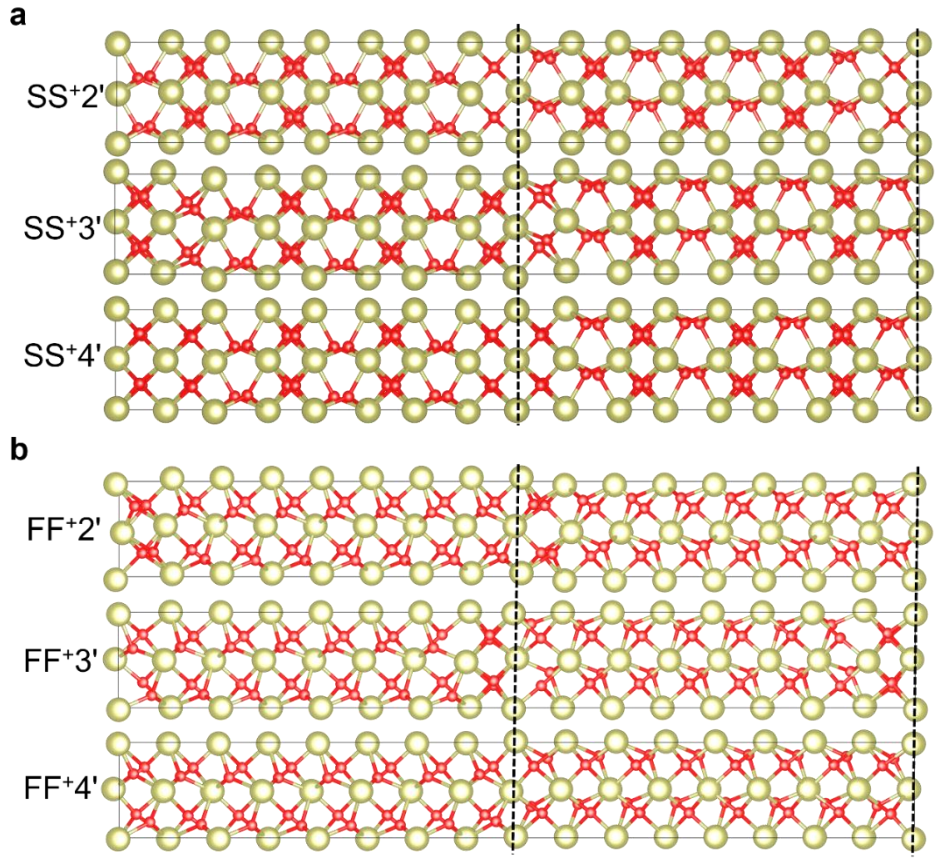

Fig. S10. The atomic structures of evenly spaced (a) SS$^+I'$ and (b) FF$^+I'$ 180° side DWs ($I$ = 2, 3, 4) in a $1 \times 8 \times 1$ periodic supercell, where $I$ corresponds to the index of their path UCP-a$I$. The DW interfaces are roughly labeled by short-dashed lines.

As illustrated in Fig. S10, the DWs originated from UCP-a$I$ ($I$ = 2, 3, 4) paths are inequivalent due to the hidden pseudo-chirality. Their atomic structures are relaxed from directly stacked u.c. building blocks, with a step-wise manner. All DW motion barriers ΔE$_{grow}$ were calculated in the transition from 3-5 u.c. spaced to 4-4 u.c. evenly spaced models.

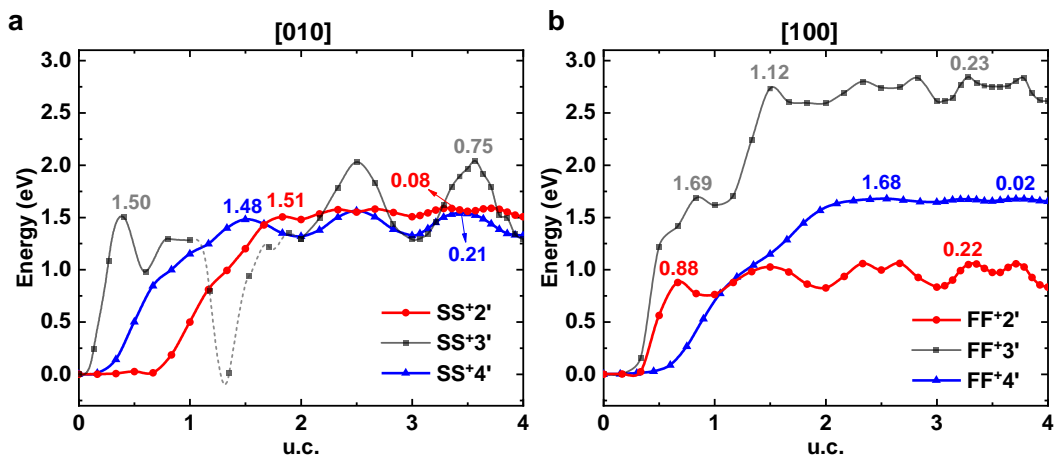

Fig. S11. Cell-by-cell nucleation-and-growth MEPs via UCP-a$I$ along (a) [010] and (b) [100] directions, respectively, calculated in $1 \times 8 \times 1$ periodic supercells. Values of representative barriers are labeled with eV units. u.c. coordinates are just for reference.

Table S2. $E_{DW}$, $\Delta E_{nuc}$, and $\Delta E_{grow}$ of UCP paths from lattice-fixed calculations. The data of UCP-b1 is extracted from proper configurations, and that of UCP-a1 is from pure OI-phase DW properly traversing T-phase transition states. The lowest values between UCP-a1 and UCP-b1, and that among UCP-a$I$ ($I$ = 2, 3, 4) are separately highlighted.

| Direction (DWs) | Ini-Fin Path | Atomic Path | $E_{DW}$ (meV/Å$^2$) | $\Delta E_{nuc}$ (eV) | $\Delta E_{grow}$ (eV) |
|---|---|---|---|---|---|
| [010] (SS$^+I'$) | {S$_1$↔S$_1$'} | UCP-b1 | **-4** | **0.47** | **0.49** |
| | | UCP-a1 | -2 | 1.22 | 1.06 |
| | {S$_1$↔S$_2$'} | UCP-a2 | 29 | 1.51 | **0.08** |
| | {S$_1$↔S$_3$'} | UCP-a3 | **25** | 1.50 | 0.75 |
| | {S$_1$↔S$_4$'} | UCP-a4 | 26 | **1.48** | 0.21 |
| [100] (FF$^+I'$) | {S$_1$↔S$_1$'} | UCP-b1 | **17** | **0.88** | **0.06** |
| | | UCP-a1 | 18 | 2.34 | 1.14 |
| | {S$_1$↔S$_2$'} | UCP-a2 | **17** | **0.88** | 0.22 |
| | {S$_1$↔S$_3$'} | UCP-a3 | 53 | 1.69 | 0.23 |
| | {S$_1$↔S$_4$'} | UCP-a4 | 33 | 1.68 | **0.03** |

To compare and get a full view among competing Hf(002) non-penetrating UCP-a paths, we have also tried performing corresponding calculations in the nucleation-and-growth scheme [5,6], same as in the main-text Fig. 3. As reflected in Fig. S11 and Table S2., UCP-a$I$ ($I$ = 2, 3, 4) paths show complex MEP diagrams during their initial switching from a single domain. Although the UCP-b1 undoubtfully wins UCP-a1, the competence among the other UCP-a$I$ seems not easy to come to a conclusion. Since they must compete in both lattice directions, [100] and [010], and in two different aspects: 1. activation energies: nucleation barriers $\Delta E_{nuc}$ and growth barriers $\Delta E_{grow}$; 2. final point stabilities: DW energies $E_{DW}$. It is a pity that the probability-dominant path among these three UCP-a$I$ paths can only be determined from further dynamical simulations.

Focusing on the current data we have, we first notice that 1-u.c. switched structures of UCP-a2 and UCP-a4 are unstable in both [100] and [010], so is the 2-u.c. switched UCP-a4 structure in [100]. They may be stabilized by using the Perdew-Burke-Ernzerhof (PBE) type generalized gradient approximation (GGA) for describing the exchange-correlation (xc) term, like in the cases of PbTiO$_3$ [6] and SS$^+$4' DW of HfO$_2$ [5]. However, it would not be meaningful to discuss results obtained from different xc-functionals, and LDA is generally considered giving good results for FE materials. Secondly, we notice the DW motions of SS$^+$3' give fluctuating MEPs during effectively switching from 1 to 2 cells (Fig. S11a). The DWs easily turn into SS$^+$1' structure (2 cells switched from S$_1$ to S$_1'$), and then the S$_1'$ cells turns into S$_3'$ cells. It is impossible under the unidirectional external electric-field. Therefore, domain nucleation via UCP-a3 can be problematic.

## Section S5. Note on the switching rate

In this section, we see how far we can go in investigating the more realistic switching rate of $Pca2_1$ FE-HfO$_2$ thin film, based on our current first-principles results.

We learn the switching dynamics of FE-HfO$_2$ thin film based on 180° DWs, according to the multiscale diffusive boundary model established by the Rappe group [3,7,8]. The polarization profile across a clean 180° (100) DW is described as:

$$P_z^{180}(x,y,z) = P_s \tanh\left(\frac{x}{\delta_x/2}\right), \tag{S1}$$

where $\delta_x$ is the domain-wall width or the diffusive coefficient, and $P_s$ is the spontaneous polarization. Its total free energy difference from a single domain ($G_0$), ignoring the elastic strain energy and strain-polarization coupling, and depolarization energies:

$$G - G_0 \approx W_l + W_g, \tag{S2}$$

consisting of the gradient energy $W_g$ and the local energy contribution $W_l$ from the Landau–Ginzburg–Devonshire phenomenological theory. The local term is written as:

$$W_l = \iiint U_{loc}(P(x,y,z;T))dxdydz, \tag{S3.1}$$

where local energy per unit cell is

$$U_{loc}(P(x,y,z;T)) = F(P(x,y,z;T)) - F(P_s(T))$$

$$= A_{loc}(T)\left(1 - \left(\frac{(P(x,y,z;T))^2}{P_s(T)}\right)^2\right) \tag{S3.2}$$

energy difference per unit cell between FE and paraelectric (PE) structures is

$$A_{loc}(T) = A_{loc}(0)\frac{P_s^4(T)}{P_s^4(0)}; \tag{S3.3}$$

And the gradient term is

$$W_g = \iiint U_m(P_z(x,y,z))dxdydz, \tag{S4.1}$$

with

$$U_m(P_z(x,y,z)) = g_m\left(\frac{\partial P_z(x,y,z)}{\partial m}\right)^2, \tag{S4.2}$$

where $g_m$ is the gradient coefficient of $P_z$ along the $m$ direction:

$$g_i = \left(\frac{3\sigma_i}{8P_s(0)}\right)^2 \frac{1}{A_{loc}(0)} = A_{loc}(0)\left(\frac{3\delta_i}{2P_s(0)}\right)^2, \tag{S4.3}$$

where $\sigma_i$ is the energy of a DW with normal vector along the $i$-axis. Note here we can also have the diffusive coefficient $\delta_i$.

Then during switching, a bulge nucleates at the DW, which would growth two-dimensionally and finally push the electric-field favored domain forward. Described as:

$$P_z(x,y,z) \approx$$
$$2P_s f^-(x,l_x,\delta_x) f^-(y,l_y,\delta_y) f^-(z,l_z,\delta_z) + P_z^{180}(x-l_x/2, y, z), \qquad (S5.1)$$

where

$$f^\pm(x,l,\delta) = \frac{1}{2}\tanh\left(\frac{x+l/2}{\delta/2}\right) \pm \frac{1}{2}\tanh\left(\frac{x-l/2}{\delta/2}\right). \qquad (S5.2)$$

The energy difference caused by the bulge is captured by the nucleation energy:
$$\Delta U_{nuc} = \Delta U_E + \Delta U_i, \qquad (S6)$$
where the polarization-field coupling term $\Delta U_E$ is

$$\Delta U_E = -E \iiint dxdydz[P_{nuc}(x,y,z) - P_{DW}(x,y,z)], \qquad (S7.1)$$

and the interfacial energy term $\Delta U_i$ is
$$\Delta U_i = \Delta W_l + \Delta W_g$$
$$= \iiint dxdydz \{[U_g(P_{nuc}) + U_{loc}(P_{nuc})] - [U_g(P_{DW}) + U_{loc}(P_{DW})]\}, \qquad (S7.2)$$

where the $P_{DW}$ and $P_{nuc}$ are the polarization profiles of a DW w/ and w/o the bulge, respectively. The critical size of the bulge can be obtained by minimizing its energy.

Now, for the switching rate (or the DW velocity) from the Merz's law

$$v(t) = v_0 \exp\left(-\frac{E_a}{E(t)}\right), \qquad (S8)$$

we can have its activation field according the Avrami theory:

$$E_{a,tot} = \frac{1}{D+1}E_{a,n} + \frac{D}{D+1}E_{a,g} \approx \frac{1}{D+1}\frac{\Delta U_{nuc}}{k_B T}E, \qquad (S8)$$

where D = 2 is the dimensionality. Here, the macroscopic switching dynamics is straightforward.

After introducing the method, we try to apply it to FE-HfO$_2$. We need the following physical parameters, ignoring the pre-exponential factor $v_0$: the energy difference between the FE and paraelectric (PE) phases $A_{loc}(T)$ (see Eq. S3.3); the bulk polarization value $P_s(T)$; the gradient coefficients along $i$ directions $g_i$ (see Eq. S4.3).

Due to the anisotropic nature of FE-HfO$_2$, we must compute the gradient coefficients separately. Available results are shown in the following Table S3, calculated based on Table S2. Note that we extract parameters from UCP-b1 proper configuration, and we use the PE energies obtained from the u.c. switching.

Table S3. The available model parameters from our first-principles data.

| Atomic Path | $P_s(0)$ (Cm$^{-2}$) | $A_{loc}(0)$ (10$^8$ Jm$^{-3}$) | $g_{x,y}$ (10$^{-11}$ m$^3$F$^{-1}$) | | $\delta_{x,y}$ (10$^{-10}$ m) | |
|---|---|---|---|---|---|---|
| | | | x | y | x | y |
| UCP-b1 | 0.68 | 9.72 | 2.32 | 0 | 2.10 | 0 |
| UCP-a2 | 0.57 | 4.11 | 7.81 | 22.73 | 4.97 | 8.48 |
| UCP-a3 | 0.57 | 9.53 | 32.75 | 7.29 | 6.68 | 3.15 |
| UCP-a4 | 0.57 | 4.11 | 29.43 | 18.27 | 9.65 | 7.60 |

Due to the stability problem in supercell models, we have not obtained domain wall energies for the 180° head-to-head and tail-to-tail DWs, and therefore we have no $g_z$ and $\delta_z$. We temporarily set $g_y$ of UCP-b1 as zero since SS$^+$1' has negative domain wall energy. The lack of $P_s(T)$ lead to absent $A_{loc}(T)$. The extraction of $P_s(T)$ with two set of UCP definitions (UCP-b1 and UCP-a*I*) can be tricky. Further studies for the multiscale model would be interesting.